\documentclass[12pt,twoside]{article}
\usepackage{fleqn,espcrc1,graphicx}

\usepackage{graphicx}
\usepackage[figuresright]{rotating}
\def\beq{\begin{equation}}
\def\eeq{\end{equation}}
\def\bea{\begin{eqnarray}}
\def\eea{\end{eqnarray}}
\def\bq{\begin{quote}}
\def\eq{\end{quote}}

\def\gappeq{\mathrel{\rlap {\raise.5ex\hbox{$>$}}
{\lower.5ex\hbox{$\sim$}}}}

\def\lappeq{\mathrel{\rlap{\raise.5ex\hbox{$<$}}
{\lower.5ex\hbox{$\sim$}}}}

\newcommand{\AmS}{{\protect\the\textfont2
  A\kern-.1667em\lower.5ex\hbox{M}\kern-.125emS}}

\title{Neutrino Physics: Theory and Phenomenology}

\author{John Ellis\address{Theoretical Physics Division, CERN \\ 
       CH 1211 Geneva 23}}%
       
\begin{document}

\maketitle

\begin{abstract}
Various issues in neutrino phenomenology are reviewed, including:
the possibility of large mixing angles in various models for neutrino
masses, difficulties for degenerate neutrinos as candidates for
hot dark matter, strategies for discriminating between different
oscillation interpretations of the atmospheric and solar neutrino
anomalies, the programme of work for long-baseline neutrino
experiments, and the possible future option of a muon storage ring
as a neutrino factory.\end{abstract}

\begin{center}
CERN-TH/99-225~~~~~~~~~~~~~~~~~hep-ph/9907458 \\
~~\\
{\it Talk at PANIC 99, XVth Particles and Nuclei International
Conference, \\
Uppsala, June 1999}
\end{center}

\section{Limits and Indications on $\nu$ Oscillations}

The discovery of neutrino masses or oscillations would take particle physics
beyond its Standard Model, and therefore requires very stringent standards of
proof and verification. Moreover, neutrino experiments are difficult, and their
history is littered with unconfirmed claims. Therefore, one must be cautious in
accepting new experimental results, and should demand that they fulfil stringent
credibility criteria. In my personal view~\cite{Helsinki}, these should
include confirmation by
more than one experiment, using more than one technique.

These criteria are obeyed by solar neutrino experiments, since five
experiments
(Homestake, Kamiokande, SAGE, GALLEX, Super-Kamiokande) see a deficit
using 3${1\over 2}$ different techniques (Cl, H$_2$0, Ga 
with two extraction schemes)~\cite{Conrad}.
Now they are also obeyed by atmospheric neutrino experiments: five experiments
(Kamiokande, IMB, Super-Kamiokande, Soudan II, MACRO) see anomalies using two
different classes of technique (H$_2$O, tracking
calorimetry)~\cite{Conrad}. Therefore, I take
these results very seriously as evidence for new physics.  
On the other hand, only one accelerator experiment (LSND) sees an
anomaly~\cite{LSND}, using
{\it a fortiori} just one technique (liquid scintillator). 
Therefore, I prefer to adopt a wait-and-see attitude to this result,
eagerly awaiting its confirmation by another experiment such as
KARMEN or MiniBooNE~\cite{Conrad}.

In the general perception, the case for atmospheric neutrino oscillations
has recently leap-frogged over that of solar neutrinos. 
This is largely because, in
addition to the sheer number of experiments reporting $\nu_\mu$ deficits, the
Super-Kamiokande Collaboration has reported dramatic effects in the
zenith-angle
distributions~\cite{SuperK}, where many systematic errors
cancel~\cite{EW}. 
Moreover, both low- and high-energy data show compatible
effects, indicating that the $\nu_\mu/\nu_e$ ratio 
decreases as $L/E$ increases, just as
expected if
$\nu_\mu$ oscillate into $\nu_\tau$ or perhaps a sterile neutrino $\nu_s$.

In the case of solar neutrinos, the overall deficit has been confirmed by
Super-Kamiokande with higher statistics~\cite{Totsuka}, but no comparable
``smoking gun" for
neutrino oscillations has yet appeared. There is a hint 
of a day-night difference~\cite{Totsuka},
but its significance remains below two standard deviations, and there is also a
hint of an distortion of the energy spectrum~\cite{Totsuka}, but a
constant suppression is still
compatible with the data at the few-percent confidence level.

Before launching into the theory of neutrino masses, it is useful to review why
the oscillation hypothesis is being pursued to the exclusion of other possible
explanations. In the case of atmospheric neutrinos, most neutrino decay
scenarios are excluded~\cite{nodecay},
flavour-changing interactions with matter are highly
disfavoured~\cite{noFCNC}, and violations
of Lorentz invariance and the Principle of Equivalence 
are disfavoured by the pattern of
zenith-angle distributions at low and high energies~\cite{LIPEOK}. In the
case of solar
neutrinos, the standard solar model is strongly supported by the
helioseismological data~\cite{helioseismo}, which do not allow substantial
changes in the solar
equation of state, and previous claims of a time dependence associated 
with the solar cycle have not been established.

\section{Neutrino Masses}

If these are non-zero, they must be much smaller than those of the corresponding
charged leptons~\cite{PDG}:
\beq
m_{\nu_e} \lappeq 2.5~{\rm eV}~, \quad\quad 
m_{\nu_\mu} \lappeq 160~{\rm keV}~, \quad\quad 
m_{\nu_\tau} \lappeq 15~{\rm eV}~,
\label{one}
\eeq
so one might think naively that they should vanish entirely. However, theorists
believe that particle masses can be strictly zero only 
if there is a corresponding
conserved charge associated with an exact gauge symmetry, which is not the case
for lepton number. Indeed, non-zero neutrino masses appear generically in Grand
Unified Theories (GUTs)~\cite{Peccei}. However, it is not necessary to
postulate new particles
to get $m_\nu \not= 0$: these could be generated by a non-renormalizable
interaction among Standard Model particles~\cite{BEG}:
\beq
{(\nu_LH)~(\nu_LH)\over M}
\label{two}
\eeq
where $M \gg m_W$ is some new, heavy mass scale. 
The most plausible guess, though,
is that this heavy mass is that of some heavy particle, perhaps a right-handed
neutrino $\nu_R$ with mass $M \sim M_{GUT}$.

In this case, one expects to find the characteristic see-saw~\cite{seesaw} 
form of neutrino mass matrix:
\beq
(\nu_L , \nu_R)~~\left(\matrix{0&m\cr
m&M}\right)~~\left(\matrix{\nu_L\cr\nu_R}\right)
\label{three}
\eeq
where the off-diagonal matrix entries 
in (\ref{three}) break SU(2) and have the form of
Dirac mass terms, so that one expects $m = 0(m_{\ell,q})$. Diagonalizing
(\ref{three}), one finds a light neutrino mass
\beq
m_\nu \simeq {m^2\over M}
\label{four}
\eeq
Choosing representative numbers $m \sim$ 10 GeV, 
$m_\nu \sim 10^{-2}$ eV one finds
$M \sim 10^{13}$ GeV, in the general ballpark of the grand unification scale.

The past year has witnessed tremendous activity in the theoretical study of
neutrino masses~\cite{vast}, of which I now pick out just a few key
features: \\
\\
{\it Other light neutrinos?}: we know from the LEP neutrino-counting constraint:
$N_\nu = 2.994 \pm 0.011$~\cite{LEPEWWG}, that any additional neutrinos
must be sterile $\nu_s$,
with no electroweak interactions or quantum numbers. 
But if so, what is to prevent
them from acquiring large masses: $m_s\nu_s\nu_s$ with $m_s \gg m_W$, as for the
$\nu_R$ discussed above? In the absence of some new theoretical superstructure,
this is an important objection to simply postulating light $\nu_s$ or $\nu_R$.\\
\\
{\it Majorana masses?}: most theorists expect the light neutrinos to be
essentially pure $\nu_L$, with only a small admixture ${\cal O}(m/M)$ of
$\nu_R$. In this
case, one expects the dominant effective neutrino mass 
term to be of Majorana type
$m_{eff}\nu_L\nu_L$, as given by (\ref{two}) or (\ref{three}). \\
\\
{\it Large mixing?}: small neutrino mixing used perhaps to be favoured, by
analogy with the Cabibbo-Kobayashi-Maskawa mixing of quarks. However, theorists
now realize that this is by no means necessary. For one thing, the off-diagonal
entries in (now considered as a 3$\times$3 matrix) (\ref{three}) need not be
$\propto m_q$ or $m_\ell$~\cite{ELLN}. Then, even if $m\propto m_\ell$, we
have no independent
evidence that mixing is small in the lepton sector. Finally, even if $m$ were to
be approximately diagonal in the same flavour basis as 
the charged leptons $e, \mu, \tau$, 
why should this also be the same case for the  heavy Majorana matrix
$M$~\cite{ELLN}?

\begin{figure}[htb]
\hglue4cm
\includegraphics[width=8.5cm]{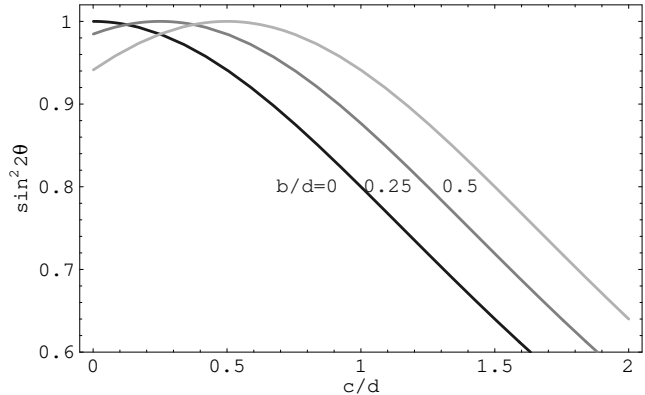}
\caption{\it Dependence of the neutrino mixing angle in the simple
two-flavour model (\ref{five}): note that one may find $\sin^2 \theta >
0.8$ for generic values of the matrix elements~\cite{ELLN}.}
\label{fig:1}
\end{figure}

\begin{figure}[htb]
\hglue4cm
\includegraphics[width=8.5cm]{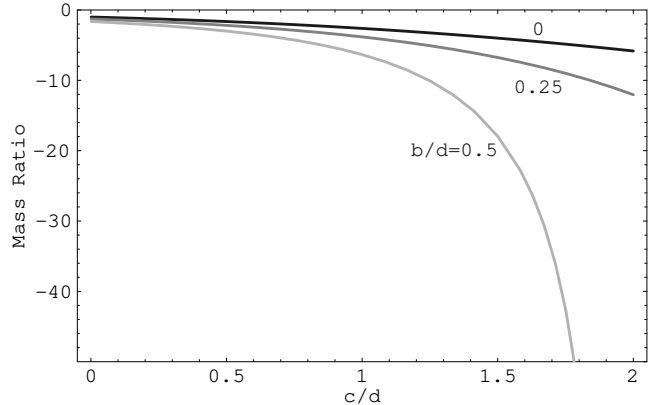}
\caption{\it Dependence of the ratio of neutrino mass
eigenvalues on the simple model (\ref{five}): note that a hierarchy
of more than an order of magnitude may be found for generic values
of the matrix elements, that may also give large $\sin^2
\theta$~\cite{ELLN}.}
\label{fig:2}
\end{figure}

Since $\sqrt{\Delta m^2_{atmo}} \sim 10^{-1}$ to $10^{-1{1\over 2}}$ eV $\gg
\sqrt{\Delta^2_{solar}} \sim 10^{-2}$ to $10^{-2{1\over 2}}$ eV
(MSW solution~\cite{MSW}) or
$10^{-5}$~eV (vacuum solution), one may ask whether large neutrino mixing is
compatible with a hierarchy of neutrino masses. To feel more comfortable about
this possibility, consider the following very simple parametrization of the
inverse of a 2$\times$2 neutrino mass matrix~\cite{ELLN}:
\beq
m^{-1}_\nu \equiv \left(\matrix{b&d\cr d&c}\right) = d \left(\matrix{b/d & 1 \cr
1 & c/d}\right)
\label{five}
\eeq
Diagonalizing this, one finds mixing:
\beq
\sin^2 2\theta = {4 d^2\over (b-c)^2 + 4d^2}
\label{six}
\eeq
which is large if $\vert d\vert \gappeq \vert b-c\vert$. However, this does not
require degeneracy of the two mass eigenvalues:
\beq
m_\pm = {2\over (b+c) \pm \sqrt{(b-c)^2+4d^2}}~,
\label{seven}
\eeq
since a large hierarchy can be obtained if $d^2 \sim bc$. 
We see in Figs.~\ref{fig:1},~\ref{fig:2}~\cite{ELLN} that
large mixing $\sin^2\theta\gappeq$ 0.8 and a hierarchy $m_+/m_- \gappeq$ 10 of
neutrino masses can be reconciled for ``reasonable" values of the dimensionless
ratios in (\ref{five}), e.g., $b/d \sim $ 0.5, $c/d \sim$ 1.5. However, it would
be difficult to accommodate the extreme hierarchy 
required by the vacuum solution
to the solar neutrino deficit in such a na\"\i ve approach.

\begin{figure}[htb]
\hglue3.5cm
\includegraphics[width=8.5cm]{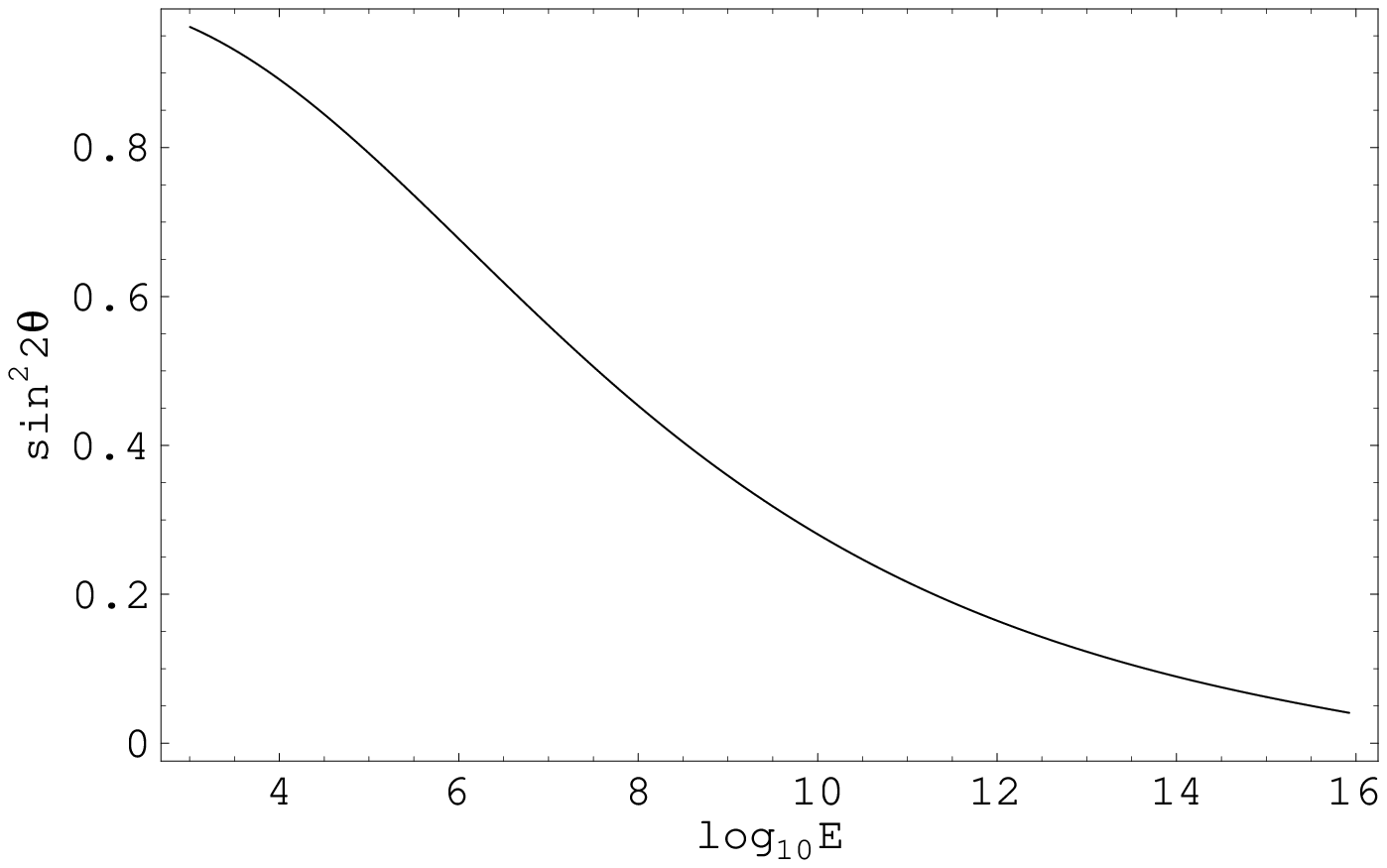}
\caption{\it Example of the possible renormalization (\ref{eight}) of the
neutrino mixing angle: note that it may be enhanced to $\sin^2 \theta >
0.8$ even if it is small at the GUT scale~\cite{ELLN}.}
\label{fig:3}
\end{figure}

There may also be significant enhancement of neutrino mixing by
renormalization-group effects between the GUT scale and the electroweak
scale~\cite{ELLN,RGE}. The
renormalization-group equation for the 2$\times$2 mixing angle $\theta$ is
\beq
16 \pi^2~{d\over dt}~(\sin^2 2\theta) = -2 (\sin^2 2\theta)~(\cos^2
2\theta)~(\lambda^2_3 - \lambda^2_2)~~{m_++m_-\over m_+ - m_-}
\label{eight}
\eeq
We see that $\theta$ can be enhanced if either the combination of Yukawa
couplings $(\lambda^2_3
- \lambda^2_2)$ is large or $(m_+-m_-)$ is small.
Fig.~\ref{fig:3}~\cite{ELLN} shows an example with
large Yukawa couplings corresponding to a large 
value of the ratio of Higgs vev's
$\tan\beta$ in a supersymmetric model. We see that a renormalization-group
enhancement of $\sin^22\theta$ from $\lappeq$ 0.2 at the GUT scale to $\gappeq$
0.9 at the electroweak scale is quite possible.

Many theoretical models of neutrino masses are circulating, often based on
specific GUT models~\cite{GUTmodels} and/or global U(1) flavour
symmetries, which illustrate some
of the points made earlier. For example, in a flipped SU(5)
model~\cite{ELLN}, the Dirac neutrino mass matrix
\beq
m^D_\nu \propto \left(\matrix{ \epsilon & {\cal O}(1) & 0 \cr 
\epsilon & {\cal O}(1) & 0 \cr
0 & 0 & {\cal O}(1) }\right)
\label{nine}
\eeq
in a first approximation, where $\epsilon$ is small, 
so that $m^D_\nu$ is not $\propto m_q$ or
$m_\ell$. There are also SO(10) models~\cite{CELW} in which entries in the
quark and lepton mass matrices
have very different U(1) weightings, so that 
lepton mixing does not parallel quark
mixing. Moreover, in U(1) models it is very 
natural to find a heavy Majorana mass
matrix that is off-diagonal in the $e,\mu,\tau$ basis. 
For example, in a 2$\times$2 model, if
the $\nu_R^{(i)}$ have U(1)  charges $n_i$, then the heavy Majorana matrix
\beq
M_{ij} \sim\epsilon^{n_i+n_j}
\label{ten}
\eeq 
where $\epsilon \ll 1$ is a U(1) hierarchy factor. Then, if $\vert n_1-n_2\vert
\ll \vert n_{1,2}\vert$, one finds
\beq
M_{ij} \propto \left(\matrix{0 & {\cal O}(1)\cr {\cal O}(1) & 0}\right)
\label{eleven}
\eeq
which is a potential source of large neutrino mixing.

In these GUT and U(1) frameworks, near-degeneracy of neutrino masses: 
$\vert m_i -
m_j\vert \ll m_{i,j}$ looks rather implausible, so that one might expect
\beq
m_3 \sim \sqrt{\Delta m^2_{atmo}} \gg m_2 \sim 
\sqrt{\Delta m^2_{solar}} \gg m_1
\label{twelve}
\eeq
However, there are also models with non-Abelian symmetries~\cite{nonAb} 
which predict
degenerate or near-degenerate neutrino masses.

Should one expect more than one large neutrino mixing angle?  This seems very
likely: for example, in the flipped SU(5) 
model~\cite{ELLN} that yields (\ref{nine}) for the Dirac
neutrino mass matrix, one also finds
\beq
M \sim \left(\matrix{ X & X & 0 \cr X & 0 & X \cr 0 & X & X}\right)
\label{thirteen}
\eeq
for the heavy Majorana mass matrix, where all the non-zero entries $X$ could be
comparable, and plausibly of order $10^{13\pm 1}$ GeV, 
as required by the see-saw
mechanism~\cite{seesaw}. The small-angle MSW solution would then appear,
possibly, to be disfavoured.

Before leaving this section, it is useful to record the general form of the
3$\times$3 neutrino mixing matrix~\cite{numix}:
\beq
\left(\matrix{\nu_e\cr\nu_\mu\cr\nu_\tau}\right) =
\left(\matrix{ c_{12}c_{13} & c_{13}s_{12} & s_{13} \cr \cr 
-c_{23}s_{12}e^{i\delta } - c_{12} s_{13}s_{23} & c_{12}c_{23} e^{i\delta} -
s_{12}s_{13}s_{23} & c_{13} s_{23} \cr\cr
s_{23}s_{12}e^{i\delta} - c_{12} c_{23} s_{13} & -c_{12} s_{23} e^{i\delta} - c_{23} s_{12}
s_{13} & c_{13} c_{23} }\right)
\left(\matrix{e^{i\alpha} & 0 & 0 \cr \cr
0 & e^{i\beta} & 0 \cr\cr 0 & 0 & 1}\right)
\left(\matrix{\nu_1\cr\cr\nu_2\cr\cr\nu_3}\right)
\label{forteen}
\eeq 
which includes two CP-violating Majorana phases 
$\alpha , \beta$ as well as three
mixing angles $\theta_{12}, \theta_{23}, \theta_{13}$ and one CP-violating phase
$\delta$ as in the quark case. Thus, a complete programme of neutrino physics
should aim at three masses, three mixing 
angles and three phases. So far, we have
experimental hints about the possible magnitudes of two mass-squared differences
$\Delta m^2$, but not the overall neutrino mass scale. One mixing angle seems to
be large: $\theta_{23} \sim 45^\circ \pm 15^\circ$ (?)~\cite{SuperK} and
one small $\theta_{13}
\sim 0^\circ \pm 20^\circ$(?)~\cite{Chooz}, but the magnitude of
$\theta_{12}$ is still
unclear, and we have no information about 
any of the phases. Indeed, the two Majorana 
phases are essentially unobservable in
experiments at energies $E \gg m_\nu$, 
though they do play a role in neutrinoless
double-$\beta$ ($\beta\beta_{0\nu})$ decay, as we discuss later.

\section{Neutrinos as Dark Matter?}

Let us set this possibility in context by first reviewing the density budget of
the Universe, in units $\Omega_? \equiv \rho_? /\rho_c$ of the critical density
$\rho_c\sim 10^{-29}$ gcm$^{-3}$. 
Generic inflation models predict $\Omega_{total}
= 1 + {\cal O}(10^{-4})$, whereas the visible baryons 
in stars, dust, etc., yield
$\Omega_{VB} \lappeq 0.01$. The success of Big-Bang Nucleosynthesis
calculations~\cite{BBN}
suggests that the overall baryon density $\Omega_B \sim 0.05$. This is not only
$\ll\Omega_{total}$ but even $\ll\Omega_m \sim 0.3$, the total mass density
inferred from observations of clusters of galaxies~\cite{clusters}.
Therefore the Universe must
contain plenty of invisible non-baryonic dark matter.

\begin{figure}[htb]
\hglue3.5cm
\includegraphics[width=8.5cm]{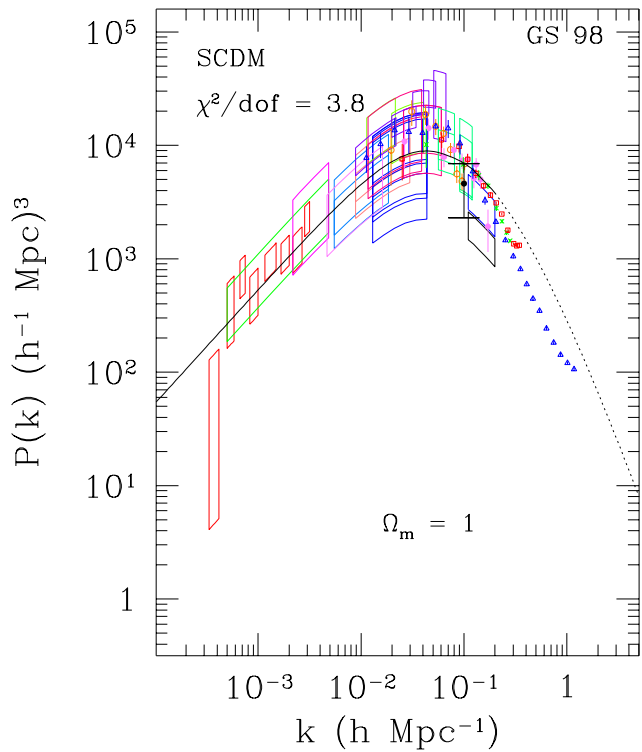}
\caption{\it Comparison of the available data on the 
power $P(k)$ in the cosmic microwave
background (parallelograms) and on large-scale structure, compared
with the standard cold dark matter model (SCDM, solid line) with
$\Omega_m = 1$: although SCDM reproduces qualitatively the trends seen
in the data, it fails at large wave number $k$~\cite{GS}.}
\label{fig:4}
\end{figure}

The astrophysical theory of structure formation suggests that most of the dark
matter is in the form of cold non-relativistic particles: $\Omega_{CDM} \gappeq
0.2$~\cite{whyCDM}. However, this theory does not fit perfectly the
combined data on large-scale
structure and the fluctuations observed in the cosmic microwave background
radiation, as seen in Fig.~\ref{fig:4}~\cite{GS}. One possibility is to
supplement cold dark matter
with hot dark matter in the form of neutrinos:
\beq
\Omega_\nu \sim \sum_\nu \left({m_\nu\over 98~{\rm ev}}\right) h^{-2}
\label{fifteen}
\eeq
where $h$ parametrizes the present Hubble expansion rate: $H \equiv $ 100 $h$ kms$^{-1}$
Mpc$^{-1}$, $h \sim 0.7 \pm 0.1$. However, alternative modifications of the
minimal cold dark matter model are possible, such as one with a cosmological
constant: $\Omega_\Lambda\sim 0.7$, which would be consistent with inflation:
$\Omega_{total}\simeq 1$, the age of the Universe, and the new data on
high-redshift supernovae~\cite{SN}.

The best one can probably say on the basis of present astrophysical and
cosmological data is that
\beq
m_\nu \lappeq 3~{\rm eV},
\label{sixteen}
\eeq
which is comparable to the direct limit (\ref{one}) on $m_{\nu_e}$. The next
generation of astrophysical and cosmological data will probably be sensitive to
$m_\nu \gappeq$ 0.3 eV~\cite{huetal}. Even $m_\nu \gappeq$ 0.03 eV may be
of cosmological
importance, but one would need to be very brave to claim astrophysical evidence
for a neutrino in the atmospheric neutrino mass range.

\begin{figure}[htb]
\hglue4cm
\includegraphics[width=8.5cm]{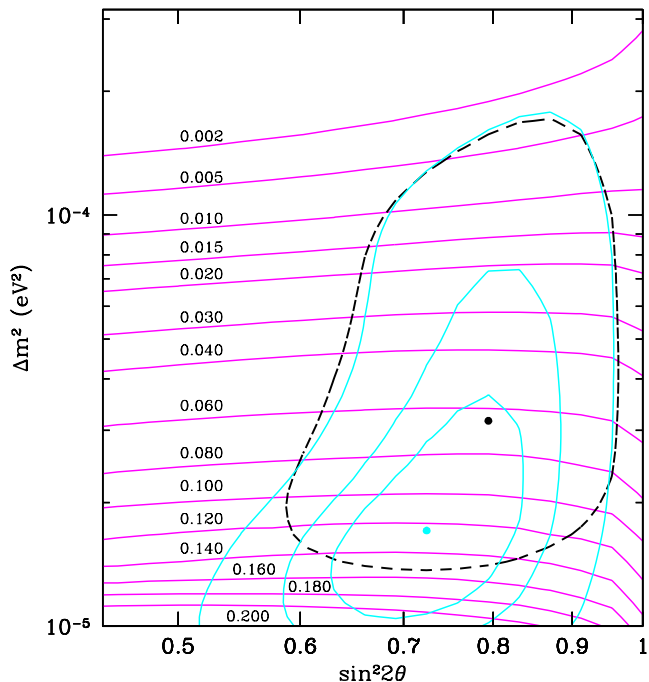}
\caption{\it Preferred region of $\sin^2 \theta$ and $\Delta m^2$
for the large-mixing-angle MSW solution to the solar neutrino
problem, both with (dashed contours) and without (grey contours)
the measured day-night asymmetry: note that $\sin^2 \theta <
0.97$~\cite{BKS}. }
\label{fig:15}
\end{figure}

Could neutrinos be degenerate, with masses $\overline{m} \gappeq$ 2 eV and close
to the direct and astrophysical limits (\ref{one}),
(\ref{sixteen})~\cite{EL}? Any such
scenario would need to respect the stringent constraint imposed by the absence of
$\beta\beta_{0_\nu}$ decay~\cite{betabeta}:
\beq
<m_\nu>_e ~\simeq~ \overline{m} ~\vert c^2_{12} c^2_{13} e^{i \alpha} + s^2_{12}c^2_{13}
e^{i\beta} + s^2_{13}\vert\lappeq 0.2~{\rm eV}
\label{seventeen}
\eeq
In view of the upper limit on $\nu_\mu - \nu_e$ mixing from the Chooz
experiment~\cite{Chooz},
let us neglect provisionally the last term in (\ref{seventeen}). In this case,
there must be a cancellation between the first two terms, requiring
$\alpha\simeq\beta + \pi$, and
\beq
c^2_{12}-s^2_{12} = \cos 2\theta_{12} \lappeq 0.1 \Rightarrow \sin^2 2\theta_{12}
\gappeq 0.99
\label{eighteen}
\eeq
Thus maximal $\nu_e-\nu_\mu$ mixing is necessary. This certainly excludes the
small-mixing-angle MSW solution and possibly even the large-mixing-angle MSW
solution, since this is not compatible with $\sin^2 2\theta = 1$ (which would
yield a constant energy-independent suppression of the solar neutrino flux), and
global fits typically indicate that $\sin^2 \theta_{12} \lappeq$ 0.97, as 
seen in Fig.~\ref{fig:15}~\cite{BKS}. Global fits before the new
Super-Kamiokande data on the energy
spectrum indicated that $\sin^2 2\theta \sim 1$ was possible for
vacuum-oscillation solutions. However, the new Super-Kamiokande analysis of the
energy spectrum now indicates~\cite{Totsuka} that, if there is any
consistent vacuum-oscillation
solution at all, it must have $\sin^2 2\theta$ considerably below 1, providing
another potential nail in the coffin of degenerate neutrinos.

\begin{figure}[htb]
\hglue3.5cm
\includegraphics[width=8.5cm]{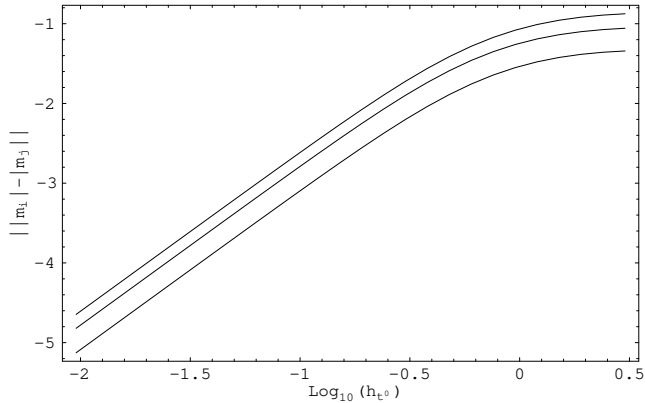}
\caption{\it Renormalization of degenerate neutrino masses as
a function of the assumed Yukawa coupling $h$: note that the degeneracy
breaking is too large, except for very small values of $h$~\cite{EL}.}
\label{fig:6}
\end{figure}

The vacuum-oscillation solution would require extreme degeneracy:
$\Delta m \sim 10^{-10} \overline{m}$, which is impossible to reconcile with a
simple calculation of neutrino mass renormalization in models with degenerate
masses at the $m_{\nu_R}$ scale~\cite{EL}, as seen in Fig.~\ref{fig:6}.
Mass-renormalization effects
also endanger the
large-angle MSW solution (which would require $\Delta m \sim 10^{-4}
\overline{m}$), and, in the context of bimaximal mixing models, also generate
unacceptable values of the neutrino mixing angles.
These renormalization problems may not be insurmountable~\cite{otherRGE},
but
they do raise
non-trivial issues that must be addressed in 
models of (near-) degenerate neutrino
masses~\cite{BRS}.

\section{How to Discriminate Between Oscillation Scenarios?}

\begin{figure}[htb]
\hglue3cm
\includegraphics[width=8.5cm]{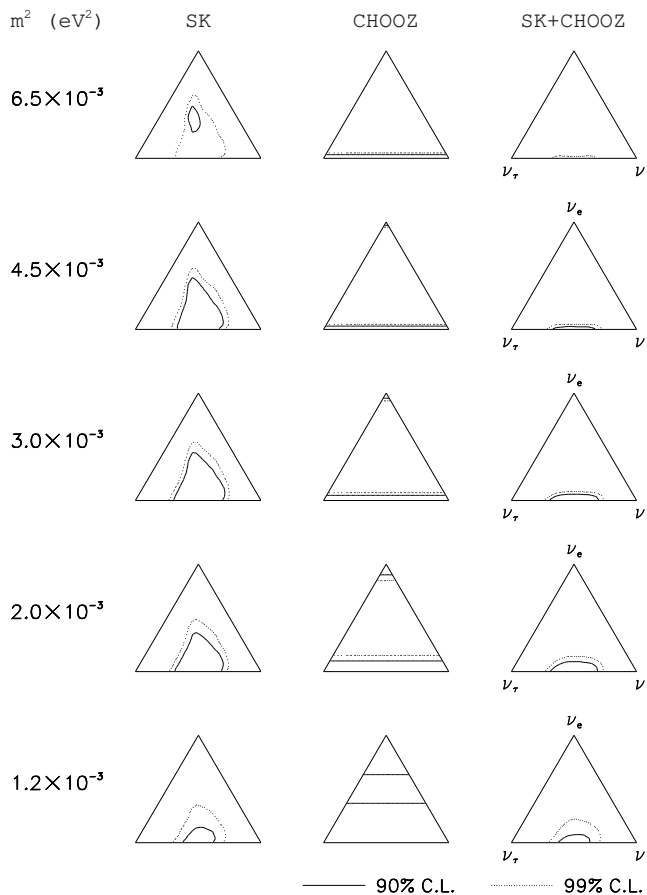}
\caption{\it Three-flavour analysis of atmospheric neutrino data:
note that a 10 \% admixture of $\nu_\mu - \nu_e$ mixing cannot be
excluded~\cite{threeflavour}.}
\label{fig:17}
\end{figure}

In the case of atmospheric neutrinos, one should consider  {\it a priori} the
possibilities of $\nu_\mu\rightarrow\nu_e , \nu_\mu\rightarrow\nu_\tau$ and
$\nu_\mu\rightarrow\nu_s$ oscillations. The first of these is certainly not
dominant, as we have learnt from the Chooz~\cite{Chooz} and
Super-Kamiokande~\cite{SuperK,Totsuka} data. However,
$\nu_\mu\rightarrow\nu_e$ oscillations could be present at a subdominant level.
Future analyses should use a complete three-flavour framework
(\ref{forteen})~\cite{threeflavour}, in
which both $\nu_\mu\rightarrow\nu_e$ and $\nu_\mu\rightarrow\nu_\tau$ 
oscillations are allowed. As seen in
Fig.~\ref{fig:17}~\cite{threeflavour},
the proportion of $\nu_\mu \to
\nu_e$ oscillations could be quite substantial, particularly for 
$3 \times 10^{-3}\, eV^2 \gappeq \Delta \, m^2 
\gappeq 1 \times 10^{-3} \, eV^2$.

Several tools to discriminate between dominant $\nu_\mu \to \nu_\tau$ and 
$\nu_\mu \to \nu_s$ oscillations are available.   One is $\pi^0$
production, which
is present in $\nu_\tau$ interactions, but absent for $\nu_\mu \to \nu_s$
oscillations.  The present data from Super-Kamiokande
yield~\cite{Totsuka}:
\beq
(\pi^0 / e)_{obs} / (\pi^0 / e)_{MC} = 1.11 \pm 0.06 \pm 0.26
\label{twenty}
\eeq
where the Monte Carlo (MC) assumes oscillations 
into neutrinos with conventional weak
interactions.  This ratio would be $\lappeq 0.7$ for $\nu_\mu \to \nu_s$
oscillations.  As seen in (\ref{twenty}), the data prefer $\nu_\mu \to \nu_\tau$
oscillations, and the statistical measurement error is relatively small, but
it is not possible to draw any definite conclusion at this
stage~\cite{Totsuka},
because of the large systematic error. This arises
from uncertainties in the $\pi^0$ production 
cross section and the detector acceptance, which should soon
be reduced by data from the nearby detector in the K2K beamline, hopefully
enabling some definitive conclusion to be drawn.

A second tool is provided by the zenith-angle distributions for atmospheric
neutrino events, which differ between $\nu_\mu \to \nu_\tau$ and $\nu_\mu \to
\nu_s$ oscillations, because of matter effects in the latter case.  As we heard
here~\cite{Totsuka}, preliminary measurements from Super-Kamiokande tend
to disfavour dominant
$\nu_\mu \to \nu_s$ at the $2-\sigma$ level, and it will be interesting to see
whether this trend is confirmed.  

In the longer run, a third tool will be provided
by the neutral-current/charged-current event ratio in long-baseline neutrino
experiments, as discussed in the next section.

In the case of solar neutrinos, there are again three main analysis tools
available to Super-Kamiokande to help discriminate between the small- and
large-angle MSW and vacuum-oscillation solutions.  One is provided by the
distortion of the energy spectrum.  
Even without including the possibility of a big $hep$
contribution~\cite{bighep}, the large-angle MSW solution is very
consistent with the latest
Super-Kamiokande data, whereas the small-angle MSW solution is somewhat
restricted, and the vacuum-oscillation solution 
appears almost excluded~\cite{Totsuka}.  This is
because the range of $\sin^2 2\theta$ and $\Delta \, m^2$ favoured by the energy
spectrum has very little overlap with that 
favoured by the overall suppression in
the rate.

The second tool is the day-night effect, 
which may also now be showing up close to
the $2-\sigma$ level~\cite{Totsuka}.  This also restricts the parameter
space of both the small-
and large-angle MSW solutions.  In the former case, a possible signature is an
enhacement as neutrinos pass through the Earth's core, which is not apparent in
the data.  No day-night effect is expected in the case of vacuum oscillations,
which may eventually turn into a problem if the current trend is confirmed.

A third tool that may soon supply some discriminating power is the seasonal
variation. In the case of the small-angle MSW solution, there should only be a
geometric effect, whereas a larger effect could appear in the other two cases,
particularly at high energies.  
Currently there is a hint of a seasonal variation
in the Super-Kamiokande data~\cite{Totsuka}, but this is not yet ready to
discriminate between the different scenarios.

In the near future, important insight into the solar-neutrino problem will be
provided by the SNO measurement of the neutral-current/charged-current
ratio. 
BOREXINO will also provide important input concerning the suppression of
intermediate-energy solar neutrinos.  Another exciting possibility is
offered by
the KamLAND experiment, which can probe the 
large-angle MSW solution directly in a
long-baseline reactor experiment.  
Within a few years, we should find a definitive
resolution of the solar neutrino problem.  In the case of atmospheric neutrinos,
this may require the input from the long-baseline 
accelerator-neutrino experiments that we now discuss.

\section{Possible Long-Baseline Accelerator Neutrino Experiments}

In the previous sections, we have reviewed the various strong pieces of evidence
for possible new neutrino physics beyond the Standard Model, which are certainly
highly indicative of neutrino masses and oscillations.  However, in the views of
many, it is necessary to use the controlled beams 
provided by accelerators - whose
fluxes, energy spectra and flavour contents 
are known and adjustable - to pin down
the interpretation of (in particular) the atmospheric-neutrino data, and to make
accurate measurements.

Two long-baseline accelerator-neutrino beams 
have already been approved.  The K2K
project extends over 250 km between KEK and the Kamioka
mine~\cite{K2K},
and has just
announced its first event in the Super-Kamiokande detector.  This will be joined
in 2002 by the 730 km NuMI project sending a beam from Fermilab to the new
MINOS~\cite{MINOS}
detector in the Soudan mine.  Under active discussion 
in Europe is the NGS project~\cite{NGS} to send a
neutrino beam from CERN to the Gran Sasso laboratory, also some 730 km distant.
This has been recommended by CERN's Scientific Policy Committee, 
and is likely to
be viewed favourably by the CERN Council if sufficient external resources can be
found.  It could start taking data in 2005.

There is a substantial programme of work for these long-baseline 
experiments.  This
includes  disappearance experiments, comparing the rates in nearby
and far detectors, as planned by K2K and MINOS.  Also important are measurements
of the neutral-current to charged-current ratio, as also planned by K2K
and MINOS.
These should provide accurate measurements of $\Delta \, m^2$ and $\sin^2 \,
2\theta$ for $\nu_\mu \to \nu_e$ or $\nu_\mu \to \nu_s$ oscillations.  The K2K
experiment is sensitive to about half of the region parameter space suggested by
Super-Kamiokande, and MINOS should cover essentially all of it.  
MINOS should also
provide some information on $\nu_e$ appearance, though it is not optimized
for $e$ detection.

In my personal view, a key measurement will be that of $\nu_\tau$
appearance via $\tau$
production.  Even if one accumulates many indirect indications that $\nu_\mu$
oscillate into $\nu_\tau$, direct proof is surely essential: ``If you have not
discovered the body, you have not proven the crime".  
Remember Jimmy Hoffa: in the
absence of a body, it was impossible to 
prove he had been murdered, let alone who
did it.  Remember also the gluon: although there were prior indirect
arguments,
everybody remembers the observation of gluon jets~\cite{threejets} as the
``discovery" of the gluon.

\begin{figure}[htb]
\hglue4cm
\includegraphics[width=8.5cm]{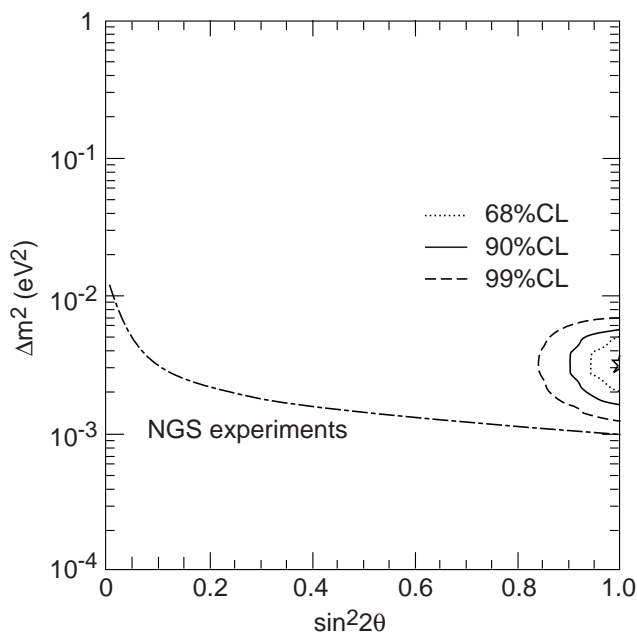}
\caption{\it Possible sensitivity of $\tau$-appearance experiments in the
proposed CERN-Gran Sasso long-baseline neutrino beam
(NGS)~\cite{NGS,tauapp}.}
\label{fig:8}
\end{figure}

The CERN-NGS beam is being optimized for $\tau$ 
production in a far detector~\cite{NGS}.  The
$\tau$ event rate $\propto \sin^2 \, 2\theta (\Delta \, m^2)^2$, and should be
${\cal O}(10)$ per year in a kiloton detector if $\Delta m^2 \sim 3 \times 10^{-3} \,
eV^2$ as suggested by the Super-Kamiokande data.  As seen in
Fig.~\ref{fig:8}~\cite{NGS}, either OPERA
or ICARUS should comfortably be able to detect $\tau$ production over all the
range of $\sin^2 \, 2\theta$ and $\Delta m^2$ indicated by Super-Kamiokande,
providing closure on the physics of atmospheric neutrinos~\cite{tauapp}.

\section{Possible Future Options}

What are the possibilities for the longer-term future?  
Accelerator options under
consideration at CERN and elsewhere include linear $e^+ e^-$
colliders - a
first generation with $\lappeq$ 1 TeV in the centre of mass~\cite{LC}, and
a
possible second
generation in the range of 2 to 5 TeV~\cite{CLIC} - a $\mu^+ \mu^-$
collider~\cite{MC,Yellow} - aiming
eventually at several TeV in the centre of mass, but with intermediate
lower-energy Higgs factory options - 
and a possible future larger hadron collider
with $\gappeq$ 100 TeV in the centre of mass.

The most relevant option for this talk may be the other physics
possibilities of an
intense $\mu$ source.  
How about stopped-$\mu$ physics with $\sim 10^{14} \mu \,
s^{-1}$?  The present limits on $\mu \to e \gamma$ and $\mu N \to eN$ could be
improved by many orders of magnitude.  
Or how about $\mu N$ scattering with $\sim$
20 GeV muons on a fixed target: 
how would this `MULFE'compare with ELFE?  Also, the rates
for  $\nu N$ scattering with a nearby (polarized?) 
target `NULFE' would be prodigious. At
CERN one could also envisage a $\mu p$ 
collider using the LHC beam.  However, the
most interesting option might be (very-)long-baseline  
neutrino physics using the
neutrinos produced  by the decays of stored muons~\cite{Geer}, which need
not be brought into
collision.  The $\mu$-decay neutrino beams are separated entirely in flavour and
charge, have a spectrum that is calculable to high precision, include equal
numbers of $\nu_\mu$ and $\nu_e$, and can easily be switched in
charge~\cite{DGH}.

We have therefore been led to propose a three-step scenario for muon storage
rings~\cite{Yellow}. The first would be a {\it $\nu$ factory}, using
$\mu$-decay neutrino beams
as the ``ultimate weapons" for $\nu$-oscillation studies.  The second step would
comprise one or more {\it Higgs factories}, 
capable of producing Higgs resonances
directly in the $s$ channel, measuring their 
total widths, restricting drastically, e.g., the MSSM parameter space,
and providing 
a new window on CP violation in the
Higgs sector: the ``ultimate weapon" for Higgs 
studies.  The third step could be a {\it multi-TeV
$\mu^+ \mu^-$ collider}. This has advantages 
over an $e^+ e^-$ collider in the same energy range,
provided by its reduced energy spread and 
its more precise energy calibration.  However, the
centre-of-mass energy may ultimately be 
limited by the neutrino-induced radiation hazard\cite{MC,King,Yellow}.

Any such programme of muon storage rings must 
face many technical problems related to the proton
driver, the target, and capturing produced pions and muons.  
In addition, muon colliders require
a large amount of beam cooling, and the $\nu$ 
radiation problem must be addressed before
progressing to a high-energy $\mu^+ \mu^-$ collider.  
However, the physics of the first-step
$\nu$ factory is already very enticing, as we now discuss.

One might envisage $10^{14} \, p$ per cycle at 
a rate of 15 Hz, producing close to $10^{21} \,
\mu^+ (\mu^-)$ per year, leading to $\nu_\mu + 
\bar{\nu}_e (\bar{\nu}_\mu + \nu_e)$ beams with
fluxes of $\sim  2 \times 10^{20}$ per year.  
These fluxes are so large that one could consider
very-long-baseline experiments with beams travelling several thousand
km~\cite{Geer,DGH,BCR,Yellow,Barger}: Fermilab to Gran
Sasso? CERN to Soudan?  either or both to Kamioka or Beijing?

\begin{figure}[htb]
\hglue1.5cm
\includegraphics[width=12cm]{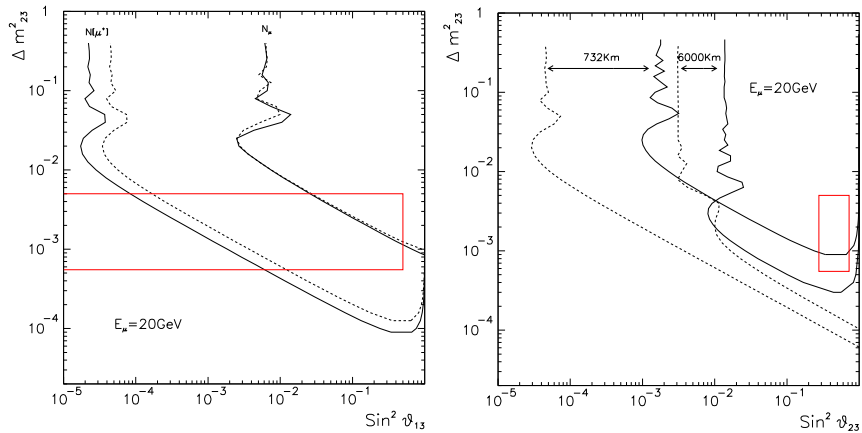}
\caption{\it The sensitivities of long-baseline neutrino experiments
using beams from a muon storage ring used as a neutrino
factory~\cite{DGH}: (a)
to search for mixing between the first- and third-generation
neutrinos via appearance (left lines) and disappearance (right lines) for
$\theta_{23} = 45^o$ (solid lines) and $30^o$ (dashed lines),
assuming a baseline of 732~km, and (b)
to search for mixing between the second- and third-generation neutrinos
via appearance (dashed lines) and disappearance (solid lines), assuming
the indicated beam lengths. The boxes represent current indications and
limits.}
\label{fig:9}
\end{figure}

\begin{figure}[htb]
\hglue4cm
\includegraphics[width=8.5cm]{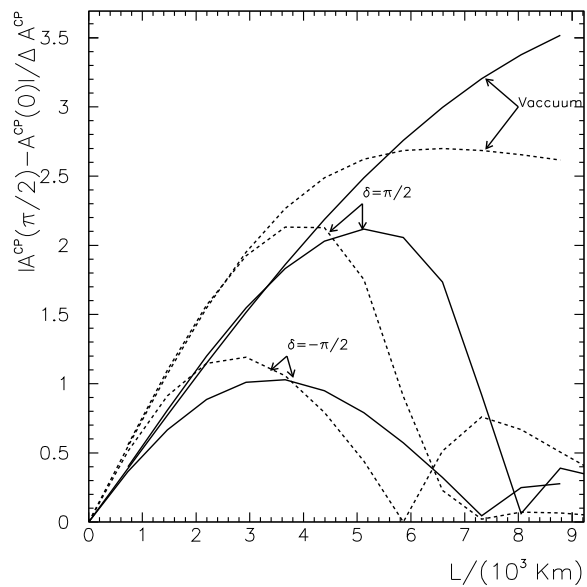}
\caption{\it Sensitivity of (very-)long-baseline experiments
to CP violating effects in neutrino oscillations~\cite{DGH}.}
\label{fig:10}
\end{figure}

The sensitivities to $\Delta m^2$ and 
$\sin^2 2\theta$ of such (very-)long-baseline experiments
have recently been studied in~\cite{DGH}.  They vary as follows with
baseline $L$ and energy $E$:
\begin{eqnarray}
& appearance & disappearance \nonumber \\
\Delta m^2 : & E_{\mu}^{-1/2} & E_{\mu}^{-1/4} L^{-1/2} \nonumber \\
\sin^2 \, 2\theta: & LE_{\mu}^{-3/2} & L^{1/2}E_{\mu}^{-3/4}
\label{twentyone}
\end{eqnarray}
As seen here and in Fig.~\ref{fig:9}, very-long-baseline experiments may
actually not confer any
benefits for appearance and disappearance studies~\cite{DGH}. 
However, the long-baseline experiments
already offer considerable improvements over the 
sensitivities of current atmospheric-neutrino
experiments.  Moreover, as seen in Fig.~\ref{fig:10}, very-long-baseline
experiments may offer a better
window on CP-violation effects in $\nu$-oscillation studies~\cite{DGH}.
Beams from $\mu$ storage rings
could be used to compare $\nu_\mu \to \nu_e$ oscillations with the $T$-reversed 
$\nu_e \to \nu_\mu$ process as well as the 
CP-conjugate process $\bar{\nu}_\mu \to \bar{\nu}_e$
(not to mention $\bar{\nu}_e \to \bar{\nu}_\mu$).  
Thus, one may begin to dream of the Holy
Grail of $\nu$-oscillation studies, 
the exploration of CP violation in the neutrino sector~\cite{Tanimoto}.
This could be connected indirectly with the baryon asymmetry of the
Universe via a leptogenesis scenario~\cite{leptogen}.
It used to be thought that neutrinos could constitute 
the dark matter: it would be ironic if
they gave birth to the visible matter.

\section{Prospects}

Neutrino physics appears finally to be 
leading particle physics beyond the straitjacket of the
Standard Model. The wealth of new data -- 
particularly from Super-Kamiokande~\cite{SuperK,Totsuka} -- is highly
suggestive
of neutrino masses and oscillations, for both 
solar and atmospheric neutrinos. In both cases,
some definitive experiments are at hand. 
In the case of solar neutrinos, these include SNO (to
see if B neutrinos have oscillated into some other flavour), BOREXINO (to see
if Be neutrinos have oscillated strongly), and KamLAND (to test the 
large-mixing-angle MSW hypothesis using the
known  flux of reactor neutrinos). Meanwhile, Super-Kamiokande is
progressing towards decisive
measurements of the spectrum distortion, 
the day-night effect and the seasonal variation of the
solar neutrino flux. In the case of atmospheric neutrinos, 
$\pi^0$ production and the
zenith-angle distribution may soon provide decisive discrimination betwen the
$\nu_\mu\rightarrow\nu_\tau$ and $\nu_\mu\rightarrow\nu_s$ 
scenarios. In this case, the definitive 
measurements will be made by long-baseline neutrino
beams from accelerators, starting with K2K. 
These have an extensive programme of work ahead of
them, including measurements of $\nu_\mu$ 
disappearance and the neutral current/charged current
ratio, as well as $\nu_e$ and $\nu_\tau$ 
appearance experiments. The detailed measurements
possible with controlled accelerator beams will 
dissipate any remaining doubts about the
interpretation of the atmospheric neutrino experiments.

In the longer run, the concept of a neutrino 
factory based on a muon storage ring offers the
prospect of a complete set of oscillation 
measurements with separated neutrino flavours and
charges, including the possibility of very-long-baseline experiments and a
quest for CP
violation. This option also offers other 
exciting opportunities in $\mu$ and $\nu$ physics, as
well as serving as a stepping-stone towards 
Higgs factories and a high-energy $\mu^+\mu^-$
collider. As never before, neutrino physics is 
entering, and perhaps diverting, the mainstream
of particle physics.

\end{document}